# Modified model for electron impact double ionization cross sections of atoms and ions

M. R. Talukder

Department of Applied Physics & Electronic Engineering, University of Rajshahi, Rajshahi-6205, Bangladesh.

E-mail: mrtalukder@ru.ac.bd

**Abastract**
A simple modification of the semiempirical model of Shevelko *et al* (J. Phys. B: At. Mol. Opt. Phys. **38**, 525 (2005)) is proposed for the calculation of electron impact double ionization cross section of $He^0$, $Li^{0,1+}$, $B^{1+}$, $C^{1+,3+}$, $O^{0-3+}$, $Ar^{0-7+}$, $Fe^{1+,3+-6+}$, $Kr^{0-4+}$, $Xe^{0-8+}$, $Pr^{1+-4+}$, $Pb^{0-9+}$, $Bi^{1+-3+,10+}$, and $U^{0,10+,13+}$ atoms and positive ions. The contributions from the direct double ionization of outer shell and indirect processes of inner shells are also considered in the proposed modification. Ionic correction and relativistic factor are also incorporated. The results of the simplified model are compared with the experimental, quantum, and other semiempirical calculations where available. It is found that the proposed modification provides better performance than those obtained by the existing semiempirical cross sections over the range of incident energies and a significant number of atomic and ionic targets considered herein for the description of experimental cross sections.

## 1. Introduction

The electron impact ionization processes of atoms and ions play important role for the understanding of collision dynamics and electron-atom interactions. Electron impact double ionization (EIDI) cross sections are comprehensively exercised in numerous practical fields for instance in plasma physics, radiation physics, ion source development, accelerator physics and elemental analysis using electron probe microanalysis (EPMA), Auger electron spectroscopy (AES), electron energy loss spectroscopy (EELS), transmission electron microscopy (TEM), and so on. Experimental measurement of EIDI cross section is extremely arduous *albeit* not impossible. Quantum calculations are complex, time-consuming and have limited applicability. Moreover, quantum calculations are based upon various approximations, the generated cross-sections are not as precise as the applied researcher desires. Experimental measurements of EIDI cross sections are usually performed at discrete incident energy points. Similarly, quantum mechanical methods generate data at particular energies and specific electronic configurations of the targets on the basis of suitable physical and mathematical approximations. In the field of applied research mentioned above, researchers require straightforward model that can provide cross-sections with adequate accuracy and computation speed at an arbitrary incident energy.

There are several quantum [1-8] and classical [9-14] models are anticipated in the last few years for the computation of EIDI cross sections. Numerous semiempirical [15-24] models are proposed to describe experimental EIDI cross sections. Out of these, models propounded by Lotz [12], Bélenger *et al* [17], and Fisher *et al* [19] have been widely used for the description of cross sections for a variety of species. These models can describe EIDI cross sections on average but unable to represent peaks found in the experimental measurements, and hence lead to a discrepancy of up to an order of 10 or more. Recently, Shevelko *et al* [15, 16] and Talukder *et al* [21, 22] proposed empirical models for the computation of EIDI cross sections for atomic and ionic species. These models can fairly describe two peaks of the EIDI cross sections as found in the experimental measurements. The models proposed by Talukder *et al* [21, 22] provide better agreement with the experimental results than those obtained by the models of Shevelko *et al* [15, 16]. Both models were developed taking into account the contribution of direct double ionization of two outer-shell electrons of the targets and also of single inner-shell ionization processes followed by autoionization with additional ejection of an electron. The models propounded by Shevelko *et al* [[15, 16] introduces, in some cases, more than 40% errors in the calculated cross sections. Moreover, they proposed two models, one for the light targets having nuclear charge $Z = 1 - 20$ [15] and the other for heavy targets of nuclear charge $Z = 21 - 92$ [16], respectively. On the other hand, the models proposed by Talukder *et al* were found to describe well the experimental EIDI cross sections for $Z = 1 - 20$ [21] and $Z = 21 - 92$ [22] both for atomic and ionic targets, respectively. But for the multi-electron targets, a significant deviation between the experimental and calculated cross sections is found mainly in the low energy region. These discrepancies are most probably due to the higher order processes including the inner-shell excitation with subsequent double auto-ionization (EDA), and the resonant excitation-triple-autoionization (RETA) due to the capture of incident electron by the target ion into a highly excited state which decays either by sequential or by simultaneous removal of three electrons. The direct and the indirect processes can be treated quantum mechanically for the estimation of EIDI cross sections with tremendous mathematical complexities taking certainly many approximations into account. But modeling codes, as used by the researchers in the applied fields, prefer analytic models providing straightforward calculation of cross-sections rather than the extensive numerical methods used in solving the problem. These facts provoked us to obtain a simple but reliable, that produce less than 20% errors, semi-empirical user-friendly single formula, in spite of two as proposed by Shevelko *et al* [15, 16] and Talukder *et al* [21, 22], for the description of plausibly accurate EIDI cross sections with atomic numbers for the extensive range from $Z = 1 - 92$ over a wide range, from $I_{th}$ to $10^5 eV$, of incident energies.

An outline of the model is sketched in Section 2. The results of the proposed model are compared with the available experimental and theoretical results and discussed in Section 3. Finally, conclusions are drawn in Section 4.

## 2. Outline of the model

The semi-empirical formula, originally developed by Shevelko *et al* [15] for multiple ionization, of Bélenger *et al* [17] is often used for double ionization by electron impact. Shevelko *et al* [15] concentrated their work for double ionization of light ions where the *K*-shell electrons are responsible for the indirect double ionization process. They calculated cross sections for the removal of *K*-shell electrons taking the Coulomb-Born approximation into account and finally presented the cross sections using a common fitting formula of Golden and Sampson [23, 24]. However, the EIDI cross section $\sigma_2$, taking into account the direct and indirect collision processes [15, 16, 19, 21, 22], for atomic and ionic systems can be written as

$$\sigma_2(E) = \sigma_{di}(E) + \sigma_{id}(E) = \sigma_{di}(E) + \sum_j \beta_j \, \sigma_j(E), \quad \text{for } E \geq I_{th} \qquad (1)$$

where $\sigma_{di}$, $\sigma_{id}$, $I_{th}$, $E$, $\beta_j$, and $\sigma_j$ are the direct double ionization cross section for simultaneous ionization of two of the outermost electrons, ionization cross section due to indirect processes related to single inner-shell ionization followed by autoionization, threshold energy for double ionization, energy of the incident electron, branching ratio for the autoionization processes, and electron impact single ionization cross section of the *j*-shell, respectively. The summation has been considered over all the subshells *j* of the initial target. It is important to know the branching ratio $\beta_j$ for autoionization of many inner-shell electrons which contribute to the double ionization cross sections. But their calculations, will be discussed later, for multi-electron system are a difficult task. $I_{th}$ for double ionization is considered as the sum of the first ($I_1$) and second ($I_2$) ionization potentials, i.e. $I_{th} = I_1 + I_2$, of the target considered.

When the incident electron is becoming closer to the *j*-shell electrons, the atom behaves as an ion of charge $q = Z - N_j$, where $N_j$ is the number of electrons in the *j*-shell. Consequently the charge cloud of electron feels attractive force towards the $j$-shell electron, thereby leading to a greater overlap of the charge clouds of the incident and target electrons and as a result EIDI cross section becomes pronounced. The effect of ion on cross section reduces with the increase of incident energy as the electron get less time in the vicinity of the atomic force field. In view of the above explanation, the ionic correction factor, which increases with the increase of charge $q$ but decreases with the incident energy of the electron, [25, 26] is modified as

$$F_i = 1 + \xi\left(\frac{q}{Zu}\right)^\alpha. \qquad (2)$$

In Eq.(2) $\xi$ and $\alpha$ are fitting parameters and $u = E/I_{th}$ is the reduced energy. It is seen from Eq.(2) that $F_i$ will be unity for $q = 0$. The values of the fitting parameters $\xi = 3.65$ and $\alpha = 1.15$ are used in the present calculation and will be discussed later.

EIDI cross section is dominated by the contribution from the inner shell ionization-autoionization process. For heavy ions, the inner shell ionization is strongly affected by resonance [27]. The relativistic effect plays a significant role in determining the strength of resonance for heavy atomic ions. Hence, for heavy targets and at high incident energies, Gryzniski's [11] relativistic factor is incorporated as a multiplying factor with Eq. (1), as applied by Deutch *et al* [18] is given by

$$F_g = \left(\frac{1+2J}{u+2J}\right)\left(\frac{u+J}{1+J}\right)^2 \left\{\frac{(1+u)(u+2J)(1+J)^2}{J^2(1+2J) + u(u+2J)(1+J)^2}\right\}^{1.5}, \qquad (3)$$

where $J = mc^2/I_{th}$, $m$ and $c$ are the mass of electron and the velocity of light in vacuum, respectively.

**2.1. Direct ionization process**

Shevelko *et al* [15] developed an empirical formula for the estimation of multiple ionization cross sections. Later on, this formula has been modified by Bélenger *et al* [17] and used for the calculation of EIDI cross sections. However, the ionization of two of the outermost electrons is related to the direct process and the indirect processes those are associated with single inner shell ionization followed by auto-ionization. Shevelko *et al* proposed two models, one for the light targets [15] having nuclear charge $Z = 3 - 26$ and the other for heavy targets [16] of nuclear charge $Z = 22 - 83$. These models are applicable for the range of incident energies $E < 50I_{th}$. On the other hand, Talukder *et al* proposed two models for the nuclear charge $Z = 1 - 18$ [21] and $Z = 21 - 92$ [22] as well. In the present work, a simple modification of Shevelko *et al* [15, 16] model is proposed for the estimation of EIDI cross sections for the direct ionization of two outer shell electrons. This straightforward modification considerably enhances not only for the wide range of nuclear charge from $Z = 1 - 92$ but also the range of incident electron energy from $I_{th}$ to $10^5 eV$. However, the modified form of the Shevelko *et al* [15] model for simultaneous ionization of two outer shell electrons is written as

$$\sigma_{di}(E) = \frac{A_{di}}{I_{th}^3} \frac{u-1}{(u+\varphi_{di})^2} \{1 - e^{-3(u-1)}\} 10^{-13} cm^2, \quad (4)$$

where $A_{di}$ and $\varphi_{di}$ are the target dependent fitting parameters of the direct ionization process. In Eq.(4), $\varphi_{di}$ has been considered as a target dependent parameter in the present work, while Shevelko *et al* [15] considered $\varphi_{di} = 0.5$ as a constant both for the light and heavy atomic and ionic systems. In Eq.(4), the dimensions of the fitting parameters $A_{di}$ and $\varphi_{di}$ are $eV^3$ and constant, respectively.

**2.2. Indirect ionization processes**

The indirect higher order processes include double auto-ionization (EDA) due to the inner-shell excitation, and the resonant excitation-triple-autoionization (RETA) due to the capture of incident electron by the target ion into a highly excited state decays by either sequential or by simultaneous removal of three electrons. Indirect double ionizations by electron impact may occur due to indirect ionization (or excitation) of an inner-shell electron through Auger decay(s), and RETA due to the capture of incident electron by target ion into a highly excited state which decays by sequential or simultaneous removal of three electrons. However, one can easily extract the contributions from the inner-shell ionization cross section to EDA processes provided that the electron impact direct single ionization cross section is estimated. The sum over single ionization cross section of Shevelko *et al* [15, 16] of the inner shell electrons leading to the net double ionization via subsequent auto-ionization processes, is modified as

$$\sigma_{id}(E) = \sum_j \beta_j \sigma_j = \sum_j \beta_j \frac{A_j}{I_j^2} \frac{x-1}{x(x+\varphi_j)} 10^{-13} cm^2, \quad (5)$$

where $x = E/I_j \geq 1$, $A_j$ and $\varphi_j$ are the fitting parameters, $I_j$ is the binding energy of the inner shell electrons. In Eq.(5), $\varphi_j$ has been considered as a target dependent parameter in the present calculation, while Shevelko *et al* [16] here also considered $\varphi_j = 5.0$ as a constant for the heavy ionic systems. The constant value of $\varphi_j = 5.0$ restricts, not only either in $Z = 3 - 26$ [15] or $Z = 22 - 83$ [16] but also for the incident electron energy $E < 50 I_{th}$, the applicability of their models for the wide range of nuclear charge $Z = 1 - 92$ that will be evidenced from the results of the proposed model. And hence $\varphi_j$ is considered as the target dependent parameter in the presented work. It is noted that the branching ratio $\beta_j$ in Eq. (5) will be considered as unity [9, 10, 15] and will be discussed later. In Eq. (5), the dimensions of the fitting parameters $A_j$ and $\varphi_j$ are $eV^2$ and constant, respectively.

Now, inserting the ionic correction factor from Eq.(2) and the relativistic effect from Eq.(3) in Eq. (1), the EIDI cross sections $\sigma_2(E)$ can finally be expressed as

$$\sigma_2(E) = F_g F_i [\sigma_{di}(E) + \sigma_{id}(E)] \, 10^{-13} cm^2. \qquad (6)$$

Eq. (6) is used for the estimation of EIDI cross sections for the range of incident electron energies from $I_{th}$ to $10^5 eV$ and for the nuclear charge from $Z = 1 - 92$.

### 3. Results and Discussions

For heavy targets, it is important to know the branching ratios for many inner-shell electrons contributing to the double ionization, as noted earlier. On the other hand, the contribution of ionization-autoionization can be described with known branching ratios for the $1s$ excitation for relatively light targets. The determinations of branching ratios for the higher shells in multi-configuration electronic systems are complicated [9, 10, 15] due to the involvement of many other interaction processes including EIDA, and REDA [27]. The ensuing multiply excited state decays especially through either sequential or simultaneous ejection of three electrons, and thus contributing to the total double ionization of the parent targets. Undoubtedly, the description of the double ionization cross-sections will be more problematic, which is not a straightforward task to solve the system either analytically or numerically and hence the branching ratios will be considered unity [28] throughout the calculation considered herein for simplicity.

The ionic parameter values $\xi = 3.65$ and $\alpha = 1.15$ for the ionic correction factor $F_i$ in Eq.(2) are optimized in such a way, where a tedious optimization process is involved, that Eq.(6) describes the best fit of EIDI cross sections with respect to the experimental data for the studied range of incident energies and for the targets considered in this work. The ionization potentials $I_1$, $I_2$, and $I_j$ used in the present calculations are collected from [29, 30] and some of them, which are not available, are calculated using Dirac-Hartree-Fock code [31]. The ionization potentials and binding energies used, and the fitting parameters obtained in the calculations are tabulated in the tables 1, 2, 3, 4 and 5. The EIDI cross sections of $He^0$, $Li^{0,1+}$, $B^{1+}$, $C^{1+,3+}$, $O^{0-3+}$, $Ar^{0-7+}$, $Fe^{1+,3+-6+}$, $Kr^{0-4+}$, $Xe^{0-8+}$, $Pr^{1+-4+}$, $Pb^{0-9+}$, $Bi^{1+-3,10+}$, and $U^{0,10+,13+}$ atomic and ionic targets are calculated and compared with the available experimental and theoretical results where available. The values of the fitting parameters of Eq. (4) and Eq.(5) are determined from the overall best fit of the predicted cross sections to the experimental data considered herein. A measure of the quality of best fit is obtained by minimizing $\chi^2$ defined by

$$\chi^2 = \sum_i \left\{ \frac{\sigma_2(E_i) - \sigma_{exp}(E_i)}{\sigma_{exp}(E_i)} \right\}^2. \qquad (7)$$

In Eq.(7), $\sigma_2(E_i)$ and $\sigma_{exp}(E_i)$ refer to the calculated and experimental cross sections at the energy point $E_i$ of the incident electron, respectively. The optimum values of the coefficients, in terms of which the parameters $A_{di}$, $\varphi_{di}$, $A_j$, $\varphi_j$, $\xi$, and $\alpha$ are defined, are obtained using a nonlinear least square curve fitting code. The parameter values are optimized in such a way that Eq.(6) describes the best EIDI cross sections with respect to the experimental data for the range of incident energies and for the targets considered in the present calculations.

The EIDI cross sections determined by the present model are depicted in figure 1 along with the experimental and theoretical results of $He^{0+}$, $Li^{0+}$, $Li^{1+}$, $B^{1+}$, $C^{1+}$, and $C^{3+}$ atomic and ionic systems. The measured data are collected from Gaudin *et al* [32], Shah *et al* [33] and theoretical data are collected from Pindzola *et al* [34], Shevelko *et al* [15] and Talukder *et al* [21] for $He^{0+}$. The cross sections of the presented model provides better agreements with the experimental measurement than those obtained by Shevelko *et al* [15] for $He^{0+}$ targets. The experimental data for $Li^{0+}$, and $Li^{1+}$ are collected from Peart *et al* [35] and theoretical data are taken from Talukder *et al* [21], respectively. It is seen from figure 1(b) and 1(c) that the EIDI cross sections of the proposed model provide better judgment with the experimental data than those presented by Talukder *et al* [21]. The experimental and theoretical data for $B^{1+}$ are taken from Scheuermann *et al* [36] and Talukder *et al* [21], respectively. Both models fairly describe the experimental results. The measured cross sections data are picked up from Westermann *et al* [37] and theoretical one are collected from Talukder *et al* [21] and Shevelko *et al* [15], respectively, for $C^{1+}$ and $C^{3+}$. The model calculations of Shevelko *et al* [15] underestimate the measured data of $C^{1+}$, while for $C^{3+}$ it fairly describe the experimental data. Hence, as a whole it is observed from figure 1 that the proposed model provides better agreement than those obtained by either Shevelko *et al* [15] or Talukder *et al* [21] with the measured cross sections.

Figure 2 compares the EIDI cross sections calculated by the proposed model along with the measured data and theoretical results for $O^{0+}$, $O^{1+}$, $O^{2+}$, and $O^{3+}$ atomic and ionic targets. The experimental data are collected for $O^{0+}$ from [38], $O^{1+}$ from [37, and references mentioned in [15]], $O^{2+}$ and $O^{3+}$ from [38, 15], respectively. The experimental measurements show that the contributions to the EIDI cross sections from the simultaneous ionization are much smaller than that from the inner-shell ionization. The EIDI cross sections for $O^{0+}$, and $O^{1+}$ show single maximum due to direct double ionization from the outer shell. The peaks in the low-energy region for $O^{2+}$ and $O^{3+}$ targets are caused by the simultaneous ionization of the outer-shell electrons having low binding energies of the outer-shell electrons, whereas in the high-energy region the contribution from the inner-shell ionization comes into play due to the high ionization potentials. The model calculations of Shevelko *et al* [15] and Talukder *et al* [21] misjudge the experimental measurements. However, the theoretical results of Shevelko *et al* [15] overestimate the experimental results from the peak to the high energy region for $O^{1+}$, while underestimate from threshold to peak region for $O^{2+}$. However, the proposed model provides better agreement with the measured data than those obtained by Shevelko *et al* [15] and Talukder *et al* [21].

The EIDI cross sections of the proposed model are displayed in figure 3 along with the theoretical and experimental results of $Ar^0$, $Ar^{1+}$, $Ar^{2+}$, $Ar^{3+}$, $Ar^{4+}$, $Ar^{5+}$ $Ar^{6+}$, and $Ar^{7+}$ atomic and ionic systems. The experimental data are collected from [32, 39-42] for $Ar^{0+}$, and from Müller *et al* [43] for $Ar^{1+}$ to $Ar^{7+}$, respectively. Theoretical calculations of Shevelko *et al* [15] underestimate the measured cross sections in low energy region, while overestimate from the peak to the high energy region of the ionic Ar systems. The reason may arise due not to take into account the ionic, as well as relativistic effects in their model. Because, the relativistic effect strongly influences the ionization-autoionization mechanisms through the resonance process [27] for the inner shell ionization that finally contributes to the double ionization cross sections and hence increases the cross sections. However, the presented model provides better agreement with the experimental data.

The EIDI cross sections determined by the present model are depicted in figure 4 along with the experimental and theoretical results of $Fe^{1+}$, $Fe^{3+}$, $Fe^{4+}$, $Fe^{5+}$, and $Fe^{6+}$ ionic systems. The measured data are collected from Stenke *et al* [44]. Theoretical data are picked up from Jha *et al* [45] for $Fe^{1+}$ and $Fe^{3+}$, and from Talukder *et al* [22] for all ionic systems. Two distinct peaks are observed in the measured data for $Fe^{5+}$, and $Fe^{6+}$ ionic targets, while the peaks of the remaining targets are indistinct. The single peak of $Fe^{1+}$, $Fe^{3+}$, and $Fe^{4+}$ targets occur due to direct ionizations of the outer shell electrons and hence the contribution of the direct part of the ionization cross sections are much greater than those of the inner shell ionization. Theoretical results of Jha *et al* [45] underestimate the measured cross sections for $Fe^{1+}$. They considered modified binary encounter model incorporating the effects of Coulombic field of the target on the incident electron taking into account Hatree-Fock velocity distribution of the target electrons. The reason of such discrepancy may arise due not to considering the indirect ionization processes involved in the collision. However, the present model provides better agreement, than those calculated by Talukder *et al* [22], with the experimental data for all Fe ionic targets over the range of incident energies considered herein.

Figure 5 compares the EIDI cross sections calculated by the proposed model along with the measured data and theoretical results for $Kr^{0+}$, $Kr^{1+}$, $Kr^{2+}$, $Kr^{3+}$, and $Kr^{4+}$. The experimental data are collected for $Kr^{0+}$ from [42, 39, 46, 47], for $Kr^{1+}$, $Kr^{2+}$ and $Kr^{3+}$ from [42], and $Kr^{4+}$ from [27, 42], respectively. The EIDI cross sections show indistinctive single maximum due to direct double ionization from the outer shells in the most cases of Kr system except for $Kr^{1+}$, and $Kr^{2+}$. The maxima of cross sections in the low-energy region are caused by the simultaneous ionization of the outer-shell electrons having their low binding energies, whereas in the high-energy region the contributions from the inner-shell ionization play significant role due to their higher ionization potentials. The open circle data [42] for $Kr^{4+}$ are the distorted wave [DW] calculation for $3d$ and $3p$ ionizations obtained using Hatree-Fock configuration-average continuum orbitals. The DW calculations are seen to be in better agreements with the experimental data in the 175-400 eV range. The relative importance of direct double ionization is seen to be small in the region between the double ionization threshold at 140.2 eV and the lowest $3d$ threshold near 150 eV. But in the high energy region, DW calculations underestimate the experimental measurements may arise due to take into account of configuration-average continuum orbitals inspite of level by level contributions and neglecting the excitation autoionization processes involved in the states considered. However, the proposed model calculations show better or at least similar agreements in most targets over the incident energies considered with the measured data and the theoretical results of Talukder *et al* [22].

The EIDI cross sections determined by the proposed model are depicted in figure 6 along with the experimental and theoretical results of $Xe^{0+}$, $Xe^{1+}$, $Xe^{2+}$, $Xe^{3+}$, $Xe^{4+}$, and $Xe^{6+}$ atomic and ionic systems. The experimental data are collected from Mathur *et al* [48], Stephan *et al* [49], and Schram *et al* [46] for $Xe^{0+}$, from Müller *et al* [50], and Achenbach *et al* [51] for $Xe^{1+}$, $Xe^{2+}$, and $Xe^{3+}$, from Müller *et al* [50, 52, 53], Pindzola *et al* [27], and Achenbach *et al* [51] for $Xe^{4+}$, and from Gregory *et al* [54] for $Xe^{6+}$, respectively. Theoretical data are picked up from Shevelko *et al* [16], and Pindzola *et al* [27] for $Xe^{4+}$, and from Talukder *et al* [22] for all atomic and ionic systems, respectively. Two separate peaks are seen for both the atomic and ionic targets of $Xe^{0+}$, $Xe^{1+}$, $Xe^{2+}$, and $Xe^{3+}$, while the peaks for ionic $Xe^{4+}$, and $Xe^{6+}$ targets are invisible in the measured data. The present calculations show better conformity with the experiment than those of the theoretical results of Talukder et al [22] for $Xe^{0+}$, $Xe^{1+}$, $Xe^{2+}$, and $Xe^{3+}$. Pindzola et al [27] estimated DW cross sections for $4d$ and $4p$ inner-shell ionizations. They obtained $4d$ cross sections taking into account Hartree-Fock term-dependent continuum orbitals, while $4p$ cross sections calculated using configuration-average orbitals. However, the DW cross sections underestimate the measured data, the reasons may arise due to considering the configuration-average orbitals instead of level by level contributions and neglecting the excitation autoionization states below the $4p$ threshold. Such states may contribute to the double ionization cross sections via excitation double autoionization. However, the present model provides better agreement, than those calculated by Talukder et al [22], with the experimental data for ionic $Xe^{4+}$, and $Xe^{6+}$ targets over the range of incident energies considered.

The EIDI cross sections obtained by the proposed model are displayed in figure 7 along with the theoretical and experimental results of $Pr^{1+}$, $Pr^{2+}$, $Pr^{3+}$, and $Pr^{4+}$ ionic systems. The experimental data are collected from Aichele et al [55], and theoretical data from Talukder et al [22], respectively, for the Pr ionic systems. Electron impact excitation double autoionization of $4d$ orbital plays significant role for all the targets. As seen from the figures, it is evident that the present calculations provide better or at least similar agreement with the experimental results.

Figure 8 compares the EIDI cross sections calculated by the proposed model along with the measured data and theoretical results for $Pb^{0+}$, $Pb^{1+}$, $Pb^{3+}$, $Pb^{4+}$, $Pb^{5+}$ and $Pb^{9+}$ atomic and ionic systems. The experimental data are, for all targets, collected from McCartney et al [56] and Fabian et al [57] and the theoretical data are collected from Talukder et al [22] except for $Pb^{0+}$. The experimental measurements show that the contributions to the EIDI cross sections from the simultaneous ionization are significant with respect to the inner-shell ionization in most targets except $Pb^{1+}$. The calculations of the proposed model provide better agreement with the measured data, for all the targets except $Pb^{1+}$, and $Pb^{2+}$, than those obtained by Shevelko et al [16] and Talukder et al [22]. The discrepancies may arise due to the differences in the experimental appearance energies with the calculated one and the existence of metastable excited ions in the measured beam [57].

The EIDI cross sections determined by the proposed model are displayed in figure 9 along with the theoretical and experimental results of $Bi^{1+}$, $Bi^{2+}$, $Bi^{3+}$, and $Bi^{10+}$ ionic systems.

The experimental data are collected from Müller *et al* [58] for $Bi^{1+}$, $Bi^{2+}$, and $Bi^{3+}$, from Scheuermann *et al* [59] for $Bi^{10+}$, and theoretical data from Talukder *et al* [22] and Shevelko *et al* [16], respectively. As seen from the figures, it is evident that the present calculations provide better or at least similar agreement with the experimental results.

Figure 10 show the experimental, theoretical and present model calculations of $U^{0+}$, $U^{10+}$, and $U^{13+}$ EIDI cross sections. The experimental data are collected from Hally *et al* [60] and Talukder *et al* [22] for $U^{0+}$, Gregory *et al* [61] for $U^{10+}$, and $U^{13+}$, respectively. Theoretical data are collected from Talukder *et al* [22] for the three targets. The proposed model describes well the experimental data over the range of incident energies and the targets considered.

## Conclusion

A semiempirical model of Shevelko *et al* [15, 16] is modified taking into account the ionic correction factor and the relativistic effect in order to calculate electron impact double ionization cross sections of atomic and ionic species over a wide range of incident electron energies from threshold to $10^5$ eV having nuclear charge from $Z = 1 - 92$. The results of the proposed model are compared with the available experimental and theoretical data. The proposed semiempirical model calculations provide better agreement, than those obtained by the existing semiempirical models, with the experimental results over the range of incident energies considered herein for the huge number of targets. Furthermore, this model can also predict cross sections for those high energies where experimental data are not available. Hence, the proposed model for its simple structure can be a prudent choice for the applied researchers where the fast generations of electron impact double ionization cross section data are required.

**References:**

[1] Becher M, Joulakian B, Le Sech C and Chrysos M 2008 *Phys. Rev. A* **77** 052710.
[2] Dürr M, Dorn A, Ullrich J, Cao S P, Czasch A, Kheifets A S, Götz J R and Briggs J S 2007 *Phys. Rev. Lett.* **98** 193201.
[3] Boeyen van R W, Watanabe N, Cooper J W, Doering J P, Moore J H and Coplan M A 2006 *Phys. Rev. A* **73** 032703.
[4] Dey R, Roy A C and Cappello C D 2006 *J. Phys. B: At. Mol. Opt. Phys*. **39** 355.
[5] Dorn A, Kheifets A, Schröter C D, Najjari B, Höhr C, Moshammer R and Ullrich J 2002 *Phys. Rev. A* **65** 032709.
[6] Lahman-Bennani A, Duprë C and Duguet A 1989 *Phys. Rev. Lett*. **36** 1582.
[7] Cappello D C, Kada I, Mansouri A and Champion C 2011 *J. Phys: Conf. Series* **288** 012004.
[8] Cappello D C, Hda H and Roy A C 1995 *Phys. Rev. A* **51** 3735.
[9] Beigman I L and V. P. Shevelko 1995 Phys. Scr. **51,** 60.
[10] Gryzinski M S and Kunc J A 1999 *J. Phys. B: At. Mol. Opt. Phys*. **32** 5789.
[11] Gryzinski M S 1965 *Phys. Rev*. **138** 336.
[12] Lotz W 1968 *Z. Phys*. **216** 241; Lotz W 1970 *Z. Phys*. **232** 101.
[13] Bell K L, Gilbody H B, Hughes J G, Kingston A E and Smith F J 1983 *J. Phys. Chem. Ref. Data* **12** 891.
[14] Tripathi D N and Rai D K 1971 *J. Chem. Phys*. **53** 1268.
[15] Shevelko V P, Tawara H, Scheuermann F, Fabian B, Müller A and Salzborn E 2005 *J. Phys. B: At. Mol. Opt. Phys*. **38** 525.
[16] Shevelko V P, Tawara H, Scheuermann F, Fabian B, Müller A and Salzborn E 2006 *J. Phys. B: At. Mol. Opt. Phys*. **39** 1499.
[17] Bélenger C, Defrance P, Salzborn E, Shevelko V P, Tawara H and Uskov D B 1997 *J. Phys. B: At. Mol. Opt. Phys*. **30** 2667.
[18] Deutsch H, Becker K and Märk T D 1996 *J. Phys. B: At. Mol. Opt. Phys*. **29** L497.
[19] Fisher V, Ralchenko Yu, Godrish A, Fisher D and Plummer E P 1995 *J. Phys. B: At. Mol. Opt. Phys*. **28** 3027.
[20] Arnaud M and Rothenflug R 1985 *Astron. Astrophys. Suppl. Ser.* **60** 425.
[21] Talukder M R, Haque A K F and Uddin M A 2009 *Euro. Phys. J. D* **53** 133.
[22] Talukder M R, Shahjahan M and Uddin M A 2012 *Phys. Scr.* **85** 015301.
[23] Golden L D and Sampson D H 1977 *J. Phys. B: At. Mol. Phys*. **10** 2229.
[24] Golden L D and Sampson D H 1980 *J. Phys. B: At. Mol. Phys*. **13** 5378.
[25] Fonts C J, Sampson D H and Zhang H L, 1999 Phys. Rev. A **59,**1329.
[26] Talukder M R 2008 *Appl. Phys. B: Lasers and Optics.* **93** 567.
[27] Pindzola M S, Griffin D C, Bottcher C, Crandall D H, Phaneuf R A, Gregory D C 1984 *Phys. Rev. A* **29** 1749.
[28] Pindzola M S 1987 *Phys. Rev. A* **35** 4548.
[29] *CRC Handbook of Chemistry and Physics* 83[rd] ed. edited by D. R. Lide 2002 (Boca Raton, London: CRC Press).
[30] Desclaux J P 1973 *At. Data Nucl. Data Tables* **12** 325.

[31] Amusia M Y and Chernysheva L V 1997 *Computations of Atomic Processes* (Bristol: Institute of Physics Publishing).
[32] Gaudin A, Hageman R, *J. Chem. Phys.* **64**, 1209 (1967).
[33] M. B. Shah, D. S. Elliot, P. McCallion and H. B. Gilbody, *J. Phys. B: At. Mol. Opt. Phys.* **21**, 2756 (1988).
[34] Pindzola M S, Robicheaux F, Colgan J 2007 *Phys Rev A* **76** 024704.
[35] Peart B, Dolder K T 1969 *J Phys B: At Mol Opt Phys* **2** 1169.
[36] Scheuermann F, Jacobi J, Salzborn E, Müller A 2001 *Giessen* (unpublished).
[37] Westermann M, Scheuermann F, Aichele K, Hathiramani D, Steidl M and Salzborn E 1999 *Phys. Scr.* **T80** 285.
[38] Ziegler D.L, Newman J.H, Smith K.A, and Stebbings R.F 1982 *Planet. & Space Sci.* **30** 451.
[39] Nagy P, Skutlartz A, Schmidt V 1980 *J. Phys. B* **13** 1249.
[40] Schram B L, de Heer F J, Van der Wiel M J, Kistemaker J 1965 *Physica* **31** 94.
[41] Van der Weil M J, El-Sherbini Th.M, Vriens L 1969 *Physica* **42** 411.
[42] Stephan K, Helm H, and Mark T.D 1980 *J. Chem. Phys.* **73** 3763.
[43] Müller A, Tinschert K, Achenbach C, Becker R, Salzborn E, 1985a *J. Phys. B* **18**, 3011.
[44] Stenke M, Hartenfeller U, Aichele K, Hathiramani D, Steidl M, and Salzborn E 1999 *J. Phys. B: At. Mol. Opt. Phys.* **32** 3641.
[45] Jha L K, Roy B N, 2006 *Eur. Phys. J. D* **37** 51.
[46] El-Sherbini Th M, Van der Wiel M J, de Heer F J, 1970 *Physica* **48** 157.
[47] Schram B L 1966 *Physica* **32** 734.
[48] Mathur D, Badrinathan 1987 *Phys Rev* A **35** 1033.
[49] Stephan K, Mark T D 1984 *J. Chem. Phys.* **81** 3116.
[50] Müller A, Achenbach C, Salzborn E and Becker R 1984 *J. Phys. B* **17** 1427.
[51] Achenbach C, Müller A, Salzborn E and Becker R 1983 *Phys. Rev. Lett.* **50** 2070.
[52] Müller A 2002 in *Experimental data for electron-ion collisions Atomic and Molecular Data and Their Applications (3rd Int. Conf. on Atomic Data and Their Applications ICAMDATA (Gatlinburg TN 24-27 April 2002)* (AIP vol 636) edited by D R Schulz, P S Krstić and F Ownby (New York: Melville).
[53] Müller A 1991 in *Physics of Ion Impact Phenomena* edited by D Mathur (Springer Ser. Chem. Phys. Vol. 54, Berlin: Springer).
[54] Gregory D C 1985 *Nucl. Instrum. & Methods* **B10/11** 87.
[55] Aichele K, Arnold W, Bräuning H, Hathiramani D, Scheuermann F, Trassel R and Salzborn E 2003 *Nucl. Instrum. Methods Phys. Res.* **B205** 437.
[56] McCartney P C E, Shah M B, Geddes J and Gilbody H B 1998 *J. Phys. B: At. Mol. Opt. Phys.* **31** 4821.
[57] Fabian B, Müller A, Brauning H, Jacobi J, Scheuermann F A and Salzborn E, 2005 *J. Phys. B: At. Mol. Opt. Phys.* **38,** 2833.
[58] Müller A, Tinschert K, Achenbach C, Salzborn E, Becker R, 1985b *Phys. Rev. Lett.* **54**, 414.
[59] Scheuermann F, Kramer K, Huber K and Salzborn E, 2004 Abstract of contributed papers 12th Int. Conf. on the Physics of Highly Charged Ions (Vilnius, Lithuania, 6-11 Sept.) ed Z Rudzikas pp **142**.

[60] Halle J C, Lo H H and Fite W 1981 *Phys. Rev. A* **23** 1708.

[61] Gregory D C, Huq M S, Meyer F W, Swenson D R, Sataka M, Chantrenne S 1990 *Phys Rev A* 41 106.

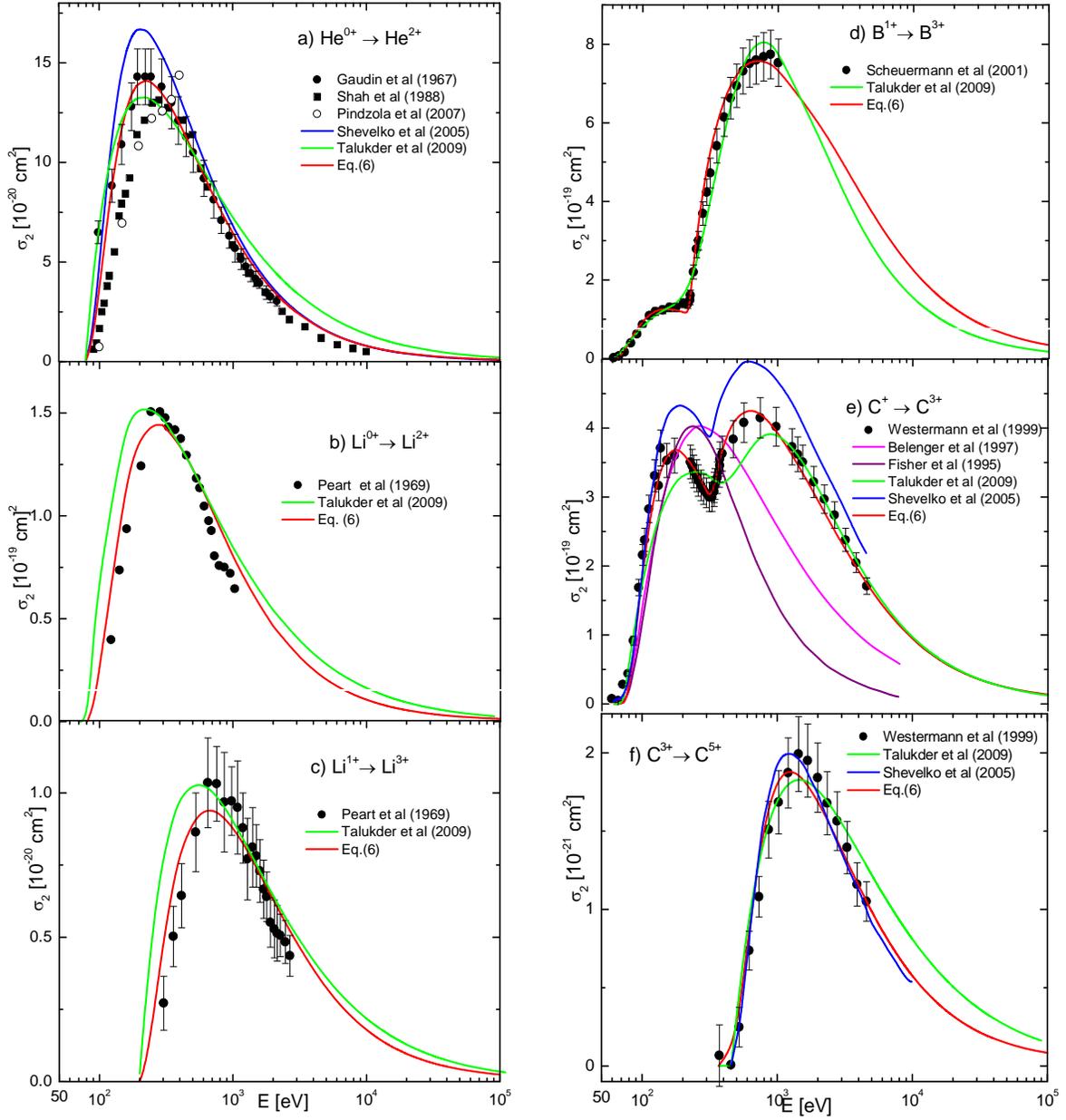

**Figure 1**. Electron impact double ionization cross sections for: a) $He^{0+}$, b) $Li^{0+}$, c) $Li^{1+}$, d) $B^{1+}$, e) $C^{1+}$ and f) $C^{3+}$.

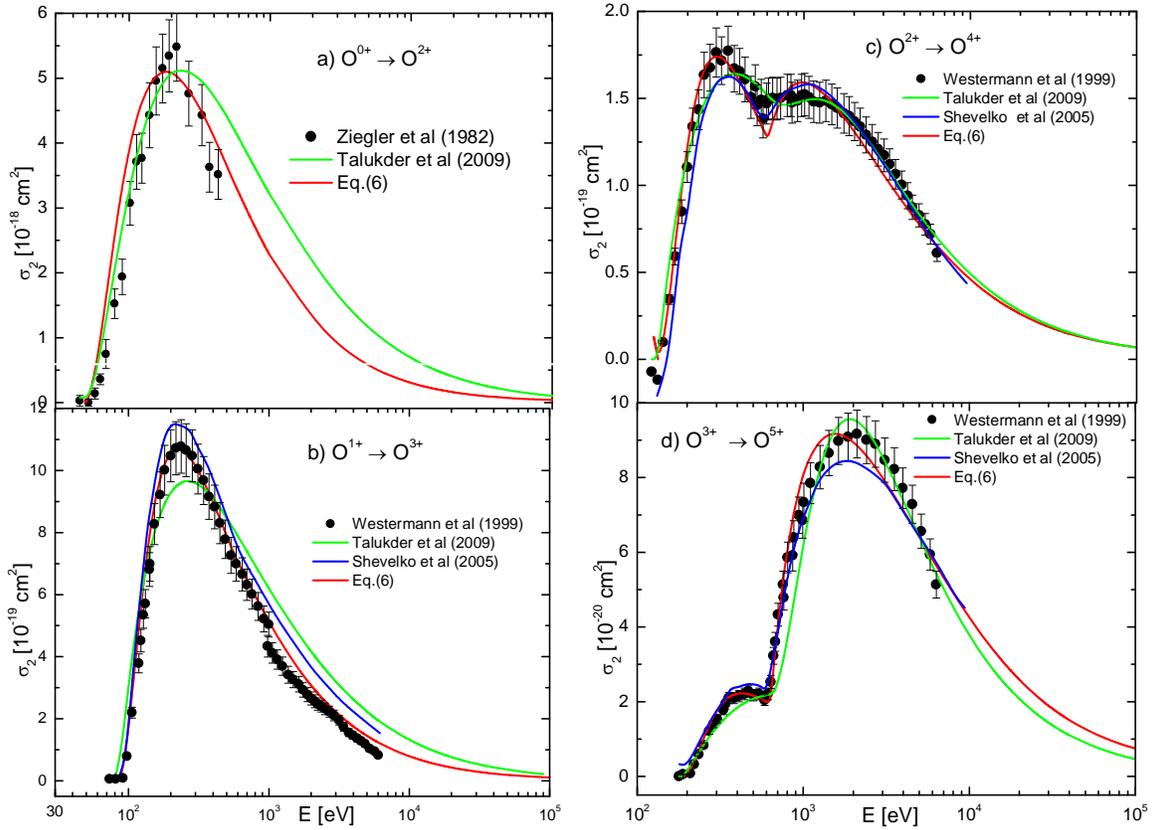

**Figure 2**. Same as in figure 1 for: a) $O^{0+}$, b) $O^{1+}$, c) $O^{2+}$, and d) $O^{3+}$.

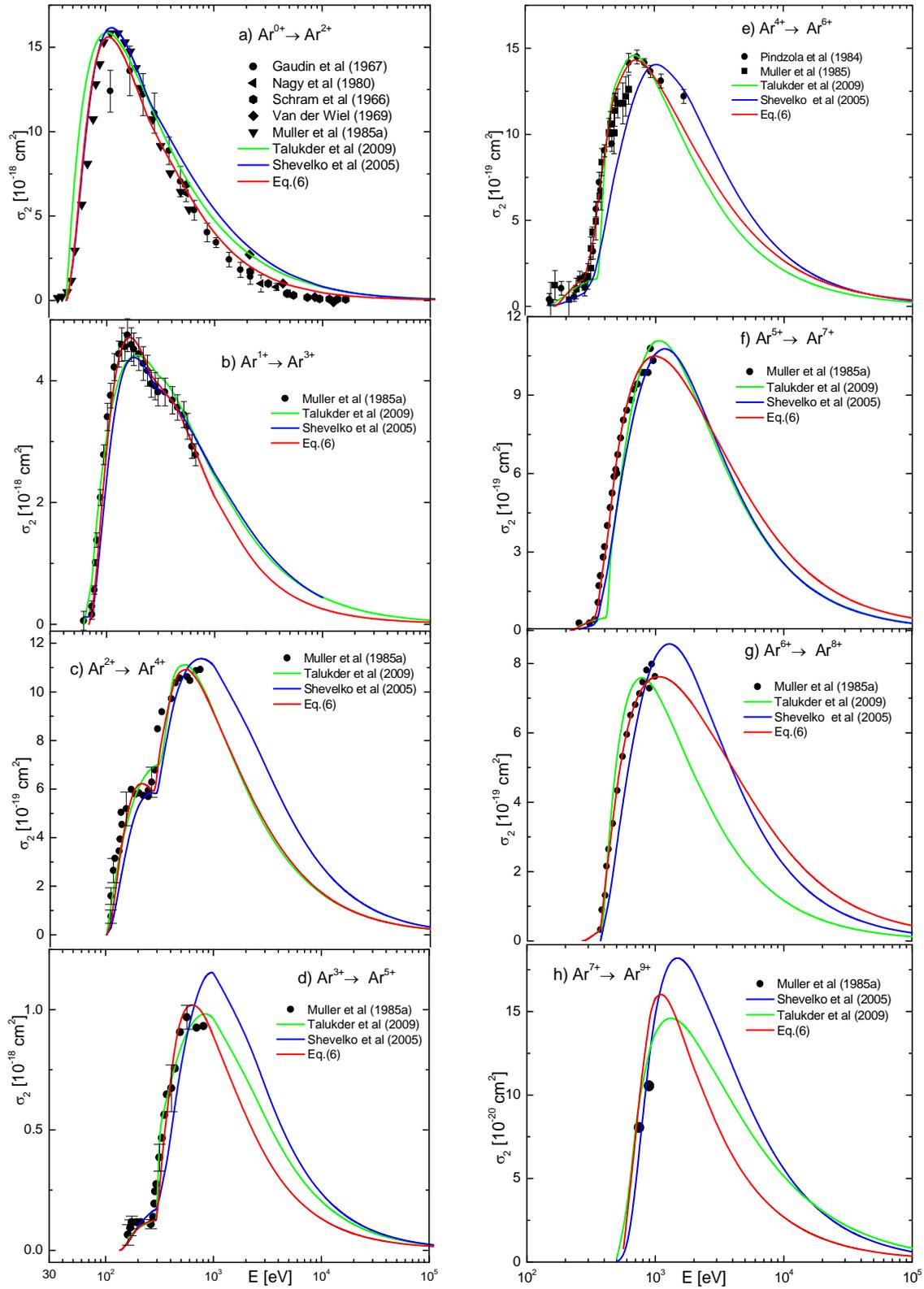

**Figure 3.** Same as in figure 1 for: a) $Ar^0$, b) $Ar^{1+}$, c) $Ar^{2+}$, d) $Ar^{3+}$, e) $Ar^{4+}$, f) $Ar^{5+}$ g) $Ar^{6+}$, and h) $Ar^{7+}$.

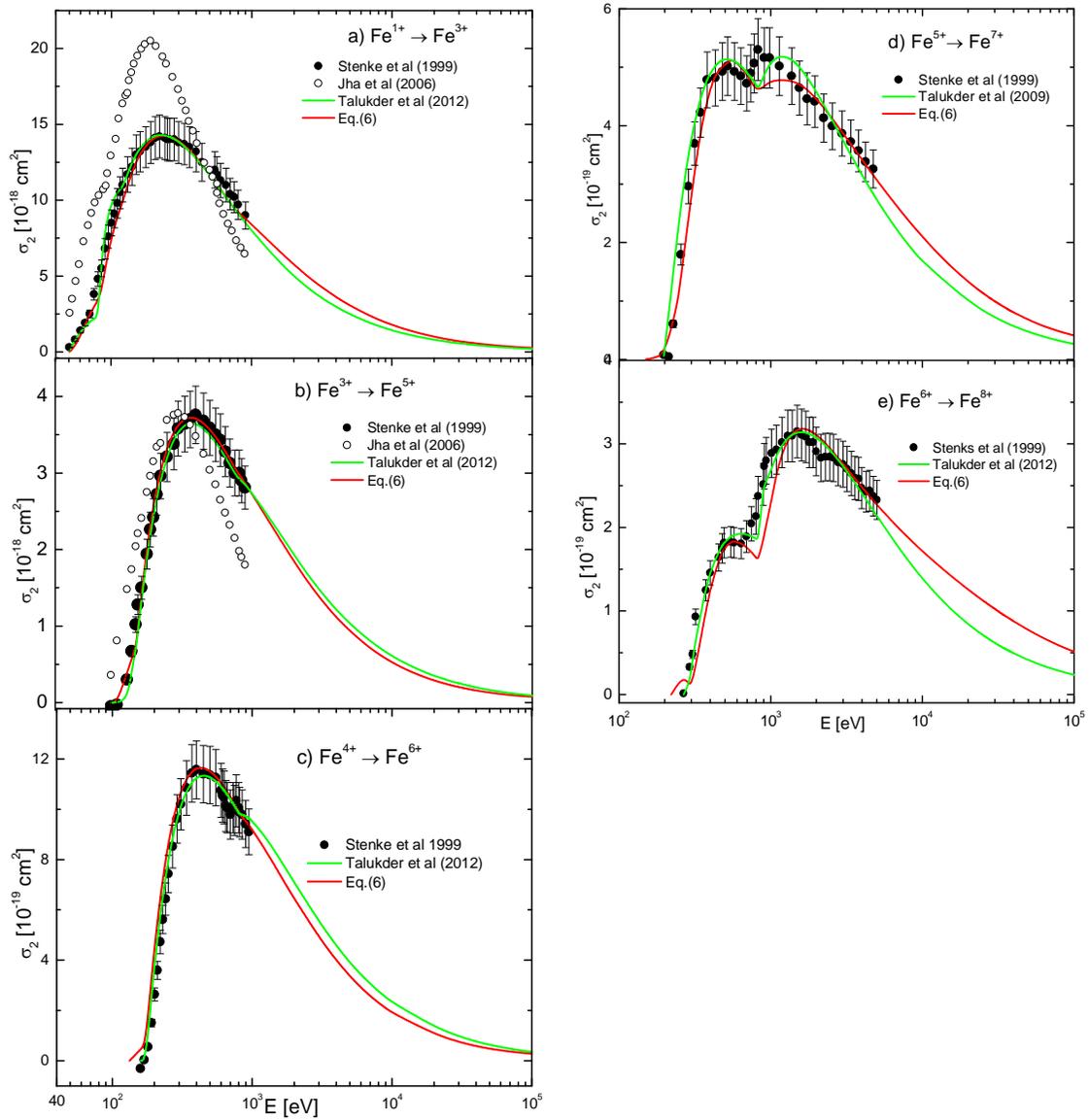

**Figure 4.** Same as in figure 1 for: a) $Fe^{1+}$, b) $Fe^{2+}$, c) $Fe^{4+}$, d) $Fe^{5+}$, and e) $Fe^{6+}$.

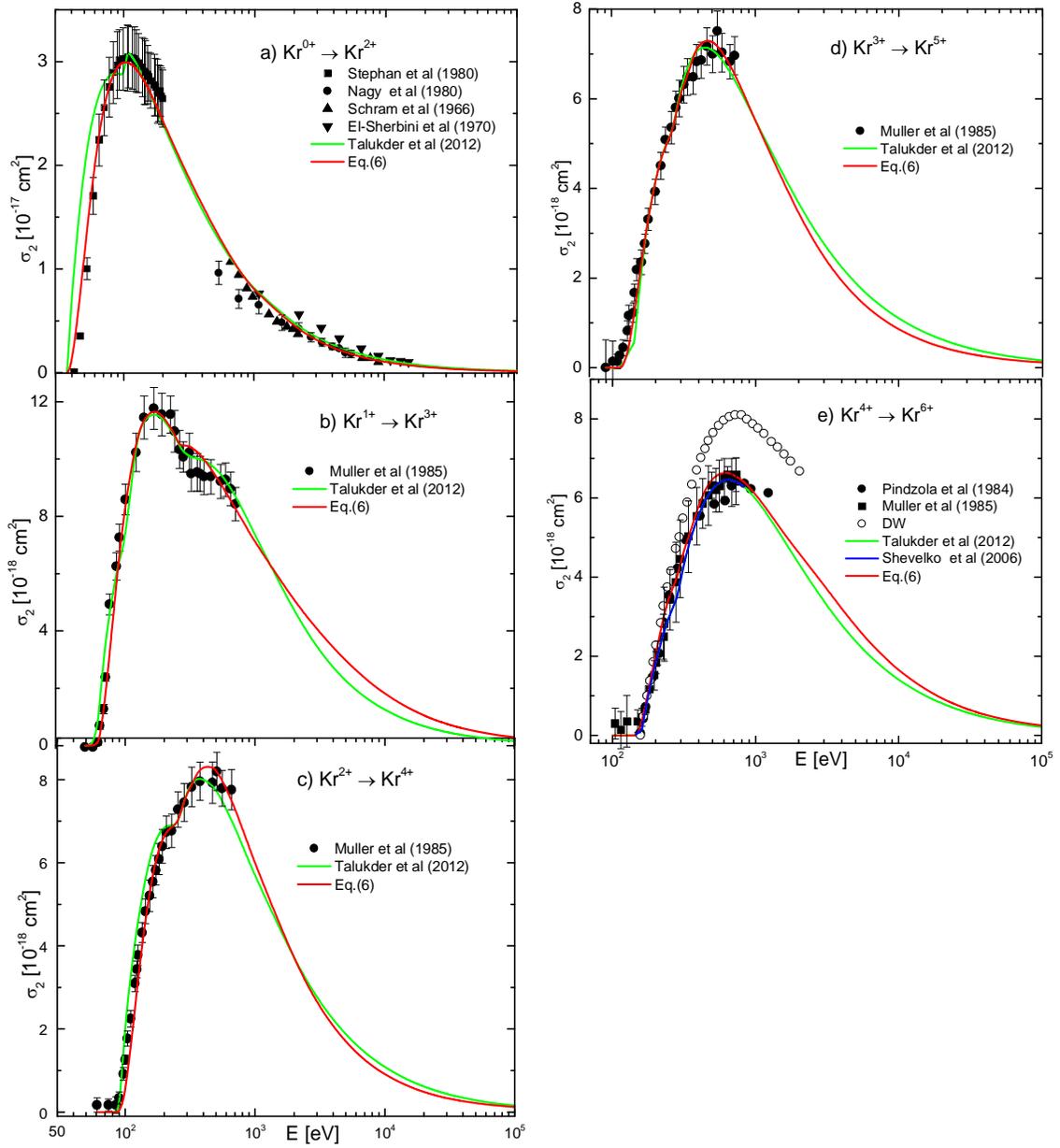

**Figure 5.** Same as in figure 1 for: a) $Kr^{0+}$, b) $Kr^{1+}$, c) $Kr^{2+}$, d) $Kr^{3+}$, and e) $Kr^{4+}$.

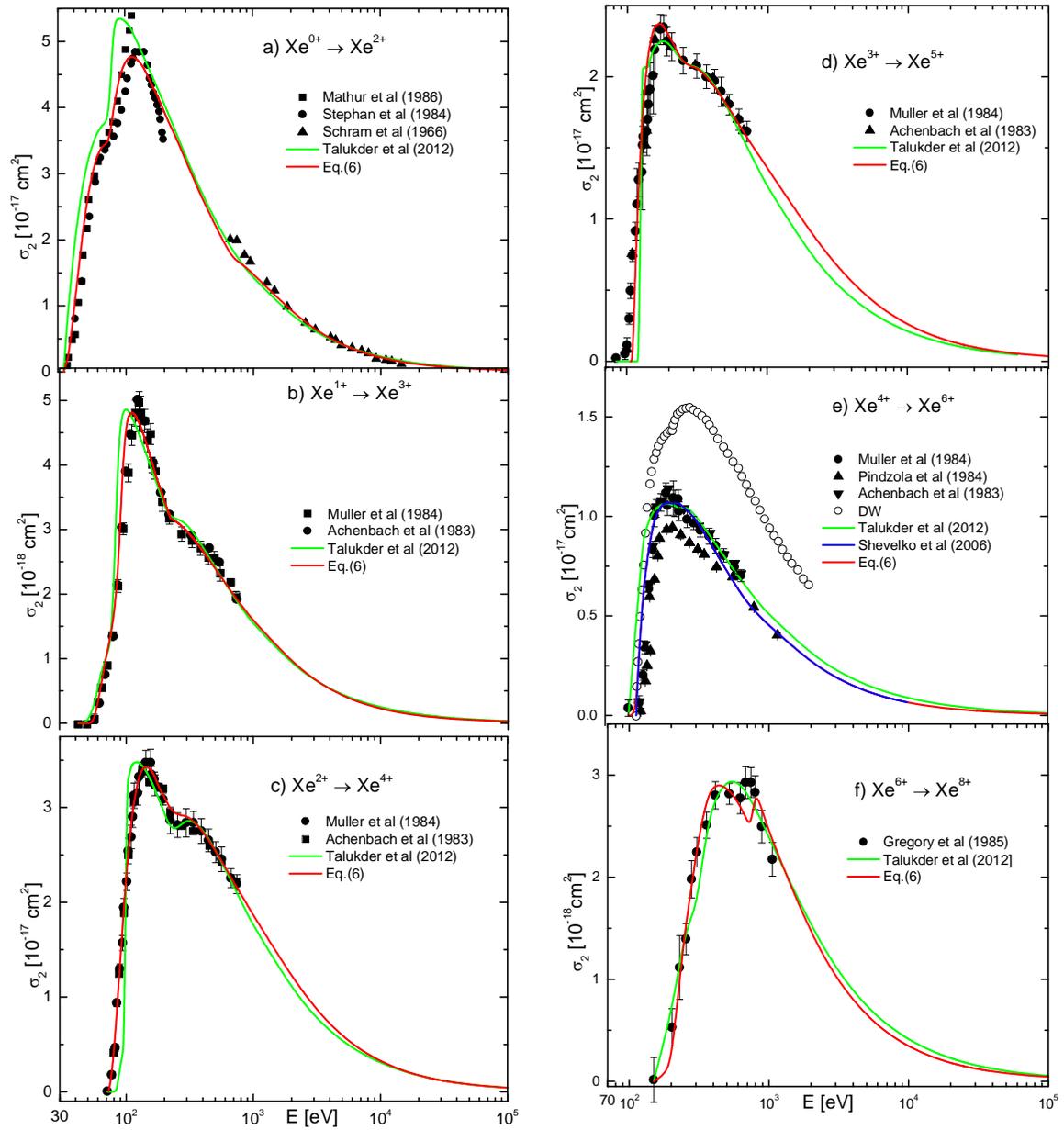

**Figure 6.** Same as in figure 1 for: a) $Xe^{0+}$, b) $Xe^{1+}$, c) $Xe^{2+}$, d) $Xe^{3+}$, e) $Xe^{4+}$, and f) $Xe^{6+}$.

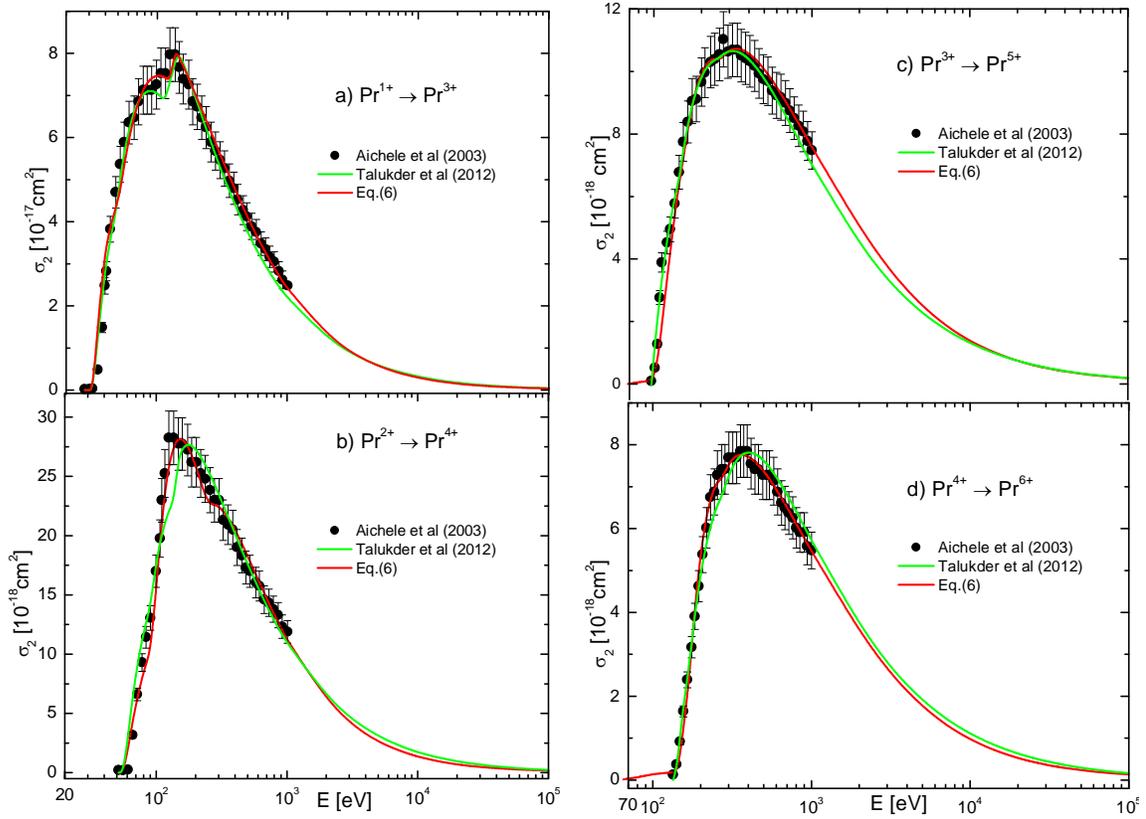

**Figure 7.** Same as in figure 1 for: a) $Pr^{1+}$, b) $Pr^{2+}$, c) $Pr^{3+}$, and d) $Pr^{4+}$.

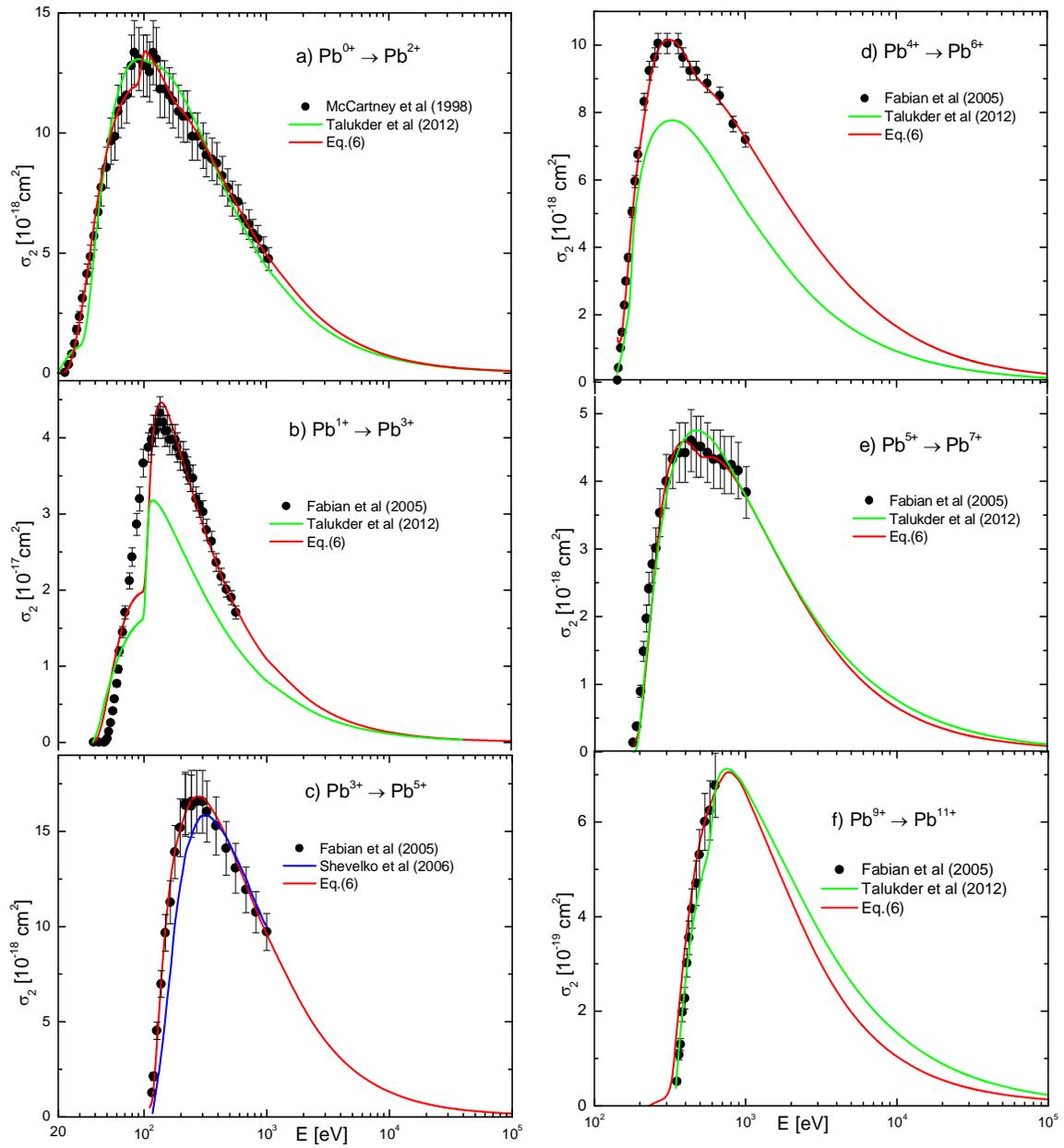

**Figure 8.** Same as in figure 1 for: a) $Pb^{0+}$, b) $Pb^{1+}$, c) $Pb^{2+}$, d) $Pb^{3+}$, e) $Pb^{4+}$, and f) $Pb^{5+}$.

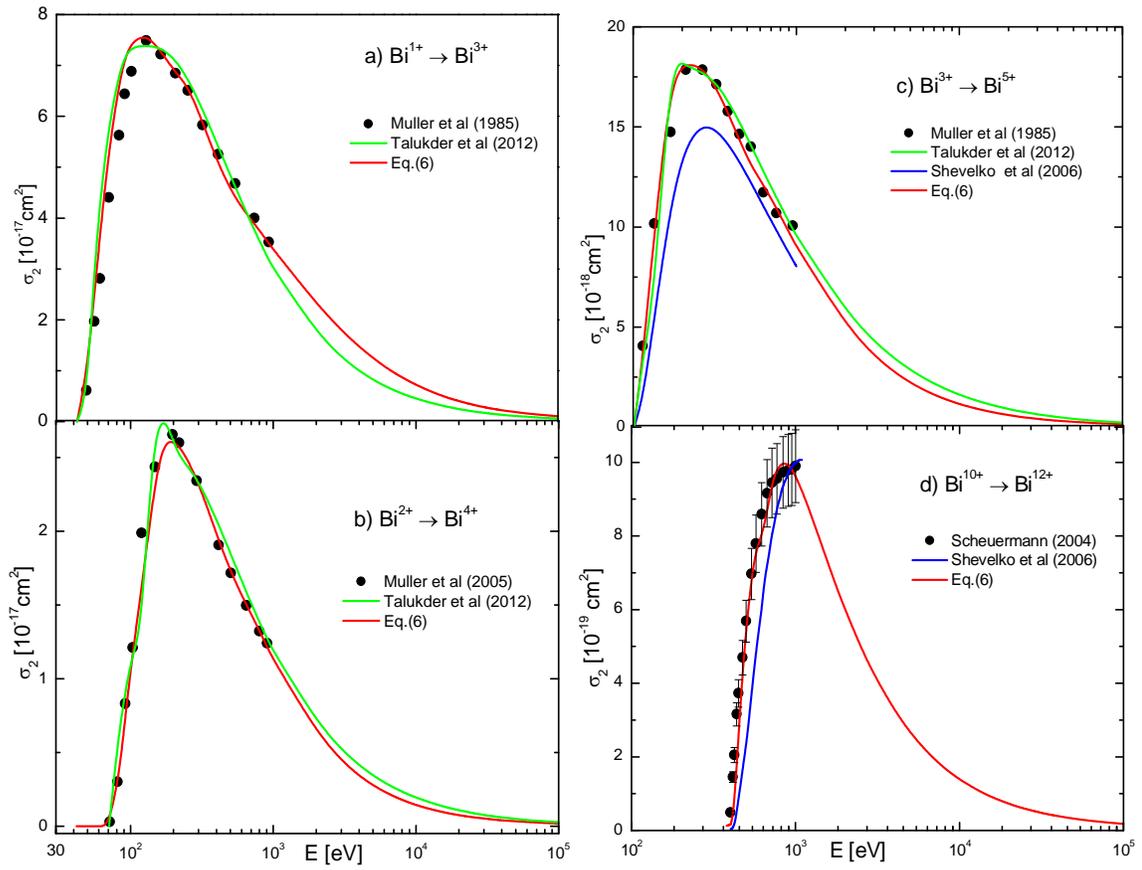

**Figure 9.** Same as in figure 1 for: a) $Bi^{1+}$, b) $Bi^{2+}$ c) $Bi^{3+}$, and d) $Bi^{10+}$.

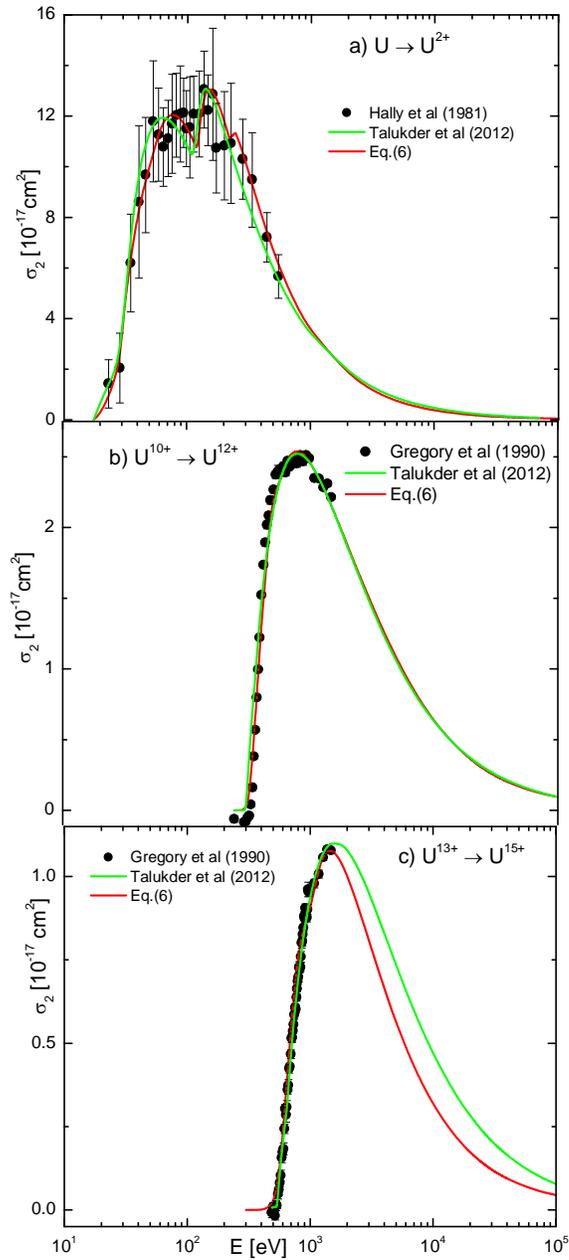

**Figure 10.** Same as in figure 1 for: a) $U^{0+}$, b) $U^{10+}$ and c) $U^{13+}$.

**Table 1.** Fitting coefficients obtained of targets considered for EIDI cross sections. The units of the fitting coefficients $A_{di}$, $A_j$, and $\varphi_j$ are $eV^3$, $eV^2$ and constant, respectively.

| Targets | $I_{th}[eV]$ $A_{di}[eV^3]$ $\varphi_{di}$ | $I_{1s}[eV]$ $A_{1s}[eV^2]$ $\varphi_{1s}$ | $I_{2p}[eV]$ $A_{2p}[eV^2]$ $\varphi_{2p}$ | $I_{2s}[eV]$ $A_{2s}[eV^2]$ $\varphi_{2s}$ |
|---|---|---|---|---|
| $He^{0+}$ | 79.00 4.90 0.75 | 24.590 0.000 0.000 | | |
| $Li^{0+}$ | 81.03 6.95 1.25 | 67.44 0.000 0.000 | | |
| $Li^{1+}$ | 198.09 7.50 2.15 | 76.01 0.000 0.000 | | |
| $B^{1+}$ | 63.08 1.60 0.50 | 139.87 5.800 10.50 | | |
| $C^{1+}$ | 72.27 7.30 0.50 | 323.87 3.000 5.00 | | |
| $C^{3+}$ | 456.57 12.50 1.75 | 366.93 0.002 5.00 | | |
| $O^0$ | 48.73 65.00 1.75 | 562.93 2.000 5.00 | – | - |
| $O^{1+}$ | 85.13 42.00 0.75 | 581.99 1.60 5.00 | – | - |
| $O^{2+}$ | 132.35 18.00 0.40 | 606.26 2.70 4.00 | – | - |
| $O^{3+}$ | 191.38 6.60 0.50 | 635.45 3.70 7.00 | – | - |

**Table 2.** Table 1 continued.

| Targets | $I_{th}[eV]$ $A_{di}[eV^3]$ $\varphi_{di}$ | $I_{1s}[eV]$ $A_{1s}[eV^2]$ $\varphi_{1s}$ | $I_{2p}[eV]$ $A_{2p}[eV^2]$ $\varphi_{2p}$ | $I_{2s}[eV]$ $A_{2s}[eV^2]$ $\varphi_{2s}$ |
|---|---|---|---|---|
| $Ar^0$ | 43.39 67.4 0.30 | - | 249.25 1.40 0.00 | 327.04 1.40 0.00 |
| $Ar^{1+}$ | 68.37 76.00 0.30 | - | 267.37 1.30 0.00 | 342.63 1.30 0.00 |
| $Ar^{2+}$ | 100.55 23.50 0.00 | - | 280.21 1.50 0.00 | 359.79 4.00 5.00 |
| $Ar^{3+}$ | 134.83 10.50 0.00 | - | 301.77 2.80 0.00 | 379.06 1.30 0.50 |
| $Ar^{4+}$ | 166.03 25.50 0.00 | - | 320.46 3.80 0.50 | 399.92 6.60 4.50 |
| $Ar^{5+}$ | 215.33 10.50 0.00 | - | 345.31 7.50 5.00 | 422.89 6.50 5.00 |
| $Ar^{6+}$ | 267.78 25.00 0.00 | - | 373.41 10.00 7.50 | 447.09 3.50 7.50 |
| $Ar^{7+}$ | 565.90 725.00 0.00 | - | 395.00 0.10 5.00 | 468.00 0.10 5.00 |

**Table 3.** Table 1 continued.

| Targets | $I_{th}[eV]$ $A_{di}[eV^3]$ $\varphi_{di}$ | $I_{3s}[eV]$ $A_{3s}[eV^2]$ $\varphi_{3s}$ | $I_{3p}[eV]$ $A_{3p}[eV^2]$ $\varphi_{3p}$ | $I_{2s}[eV]$ $A_{2s}[eV^2]$ $\varphi_{2s}$ | $I_{2p}[eV]$ $A_{2p}[eV^2]$ $\varphi_{2p}$ |
|---|---|---|---|---|---|
| $Fe^{1+}$ | 49.60 25.30 0.20 | 122.22 4.90 5.00 | 83.41 5.30 3.00 | 878.25 1.30 5.00 | 755.20 71.30 5.00 |
| $Fe^{3+}$ | 128.95 25.30 0.20 | 151.07 3.30 2.00 | 111.92 2.30 7.00 | 908.58 0.50 2.00 | 785.52 11.30 5.00 |
| $Fe^{4+}$ | 173.92 25.30 0.20 | 172.50 2.00 2.50 | 132.78 0.30 7.00 | 932.49 0.50 2.00 | 809.34 5.30 5.00 |
| $Fe^{6+}$ | 286.00 130.30 0.00 | 221.06 0.10 2.00 | 180.10 0.30 7.00 | 988.60 42.50 25.00 | 847.20 2.90 0.00 |
| $Kr^{0+}$ | 36.95 103.30 0.70 | 295.32 3.30 2.00 | 226.78 1.30 0.00 | 981.01 0.50 2.00 | 855.80 11.30 5.00 |
| $Kr^{1+}$ | 61.31 181.30 0.70 | 268.71 39.30 10.00 | 203.56 0.10 0.00 | 1774.74 0.50 2.00 | 1593.85 61.30 5.00 |
| $Kr^{3+}$ | 117.20 553.30 0.70 | 335.61 9.30 2.00 | 266.89 4.30 0.00 | 1943.53 0.50 2.00 | 1755.89 11.30 5.00 |
| $Kr^{4+}$ | 143.20 803.30 0.70 | 351.66 9.30 2.00 | 282.87 34.30 12.00 | 1959.95 0.50 2.00 | 1772.32 21.30 0.00 |

**Table 4.** Table 1 continued.

| Targets | $I_{th}[eV]$ $A_{di}[eV^3]$ $\varphi_{di}$ | $I_{1n}[eV]$ $C_{1n}[eV^2]$ $\varphi_{1n}$ | $I_{2n}[eV]$ $C_{2n}[eV^2]$ $\varphi_{2n}$ | $I_{3n}[eV]$ $C_{3n}[eV^2]$ $\varphi_{3n}$ | $I_{4n}[eV]$ $C_{4n}[eV^2]$ $\varphi_{4n}$ |
|---|---|---|---|---|---|
| Xe | 33.34 / 53.30 / 0.00 | 75.61(5d) / 4.60(5d) / 0.00(5d) | 163.55(4p) / 1.50(4p) / 0.00(4p) | 213.85(4s) / 6.90(4s) / 0.00(4s) | 710.96(3d) / 51.30(3d) / 0.00(3d) |
| $Xe^{1+}$ | 53.33 / 100.30 / -0.20 | 85.59(5d) / 3.50(5d) / -0.90(5d) | 173.53(4p) / 1.50(4p) / 0.00(4p) | 223.86(4s) / 30.90(4s) / 2.00(4s) | 721.02(3d) / 5.30(3d) / 0.00(3d) |
| $Xe^{2+}$ | 72.68 / 370.00 / -0.30 | 96.69(5d) / 0.60(5d) / 0.00(5d) | 184.63(4p) / 2.00(4p) / -0.50(4p) | 235.01(4s) / 58.90(4s) / 4.00(4s) | 732.29(3d) / 5.30(3d) / 0.00(3d) |
| $Xe^{3+}$ | 105.84 / 85.00 / -0.70 | 108.76(5d) / 5.00(5d) / -0.70(5d) | 196.71(4p) / 1.50(4p) / -0.50(4p) | 247.14(4s) / 52.00(4s) / 5.00(4s) | 744.61(3d) / 5.30(3d) / 0.00(3d) |
| $Xe^{4+}$ | 116.64 / 15.00 / -1.00 | 121.68(5d) / 4.00(5d) / -0.40(5d) | 209.64(4p) / 1.00(4p) / -0.50(4p) | 260.14(4s) / 5.90(4s) / 5.00(4s) | 757.88(3d) / 5.30(3d) / 0.00(3d) |
| $Xe^{6+}$ | 187.74 / 365.00 / -0.30 | 149.80(5d) / 0.10(5d) / 0.00(5d) | 237.82(4p) / 0.10(4p) / -0.30(4p) | 288.48(4s) / 6.00(4s) / 2.00(4s) | 787.02(3d) / 2.30(3d) / -1.00(3d) |
| $Pr^{1+}$ | 48.24 / 55.00 / 0.50 | 32.29(5p) / 10.50(5p) / 5.00(5p) | 50.61(5s) / 1.60(5s) / -0.30(5s) | 136.16(5d) / 1.60(5d) / -1.00(5d) | 243.51(4p) / -0.01(4p) / -0.30(4p) |
| $Pr^{2+}$ | 89.35 / 205.00 / -0.60 | 38.77(5p) / 3.00(5p) / 2.00(5p) | 57.07(5s) / 0.01(5s) / 5.00(5s) | 142.77(5d) / 0.05(5d) / -1.00(5d) | 250.11(4p) / 14.00(4p) / 0.30(4p) |
| $Pr^{3+}$ | 96.51 / 305.00 / 0.00 | 50.58(5p) / 0.01(5p) / -0.30(5p) | 68.34(5s) / 0.10(5s) / 5.00(5s) | 156.02(5d) / 7.70(5d) / 5.00(5d) | 263.30(4p) / 17.00(4p) / 5.00(4p) |
| $Pr^{4+}$ | 138.93 / 255.00 / -0.50 | 62.99(5p) / 0.01(5p) / -0.30(5p) | 80.16(5s) / 0.10(5s) / 5.00(5s) | 170.16(5d) / 4.90(5d) / 1.00(5d) | 277.39(4p) / 17.00(4p) / 5.00(4p) |

**Table 5.** Table 1 continued.

| Targets | $I_{th}[eV]$ $A_{di}[eV^3]$ $\varphi_{di}$ | $I_{1n}[eV]$ $C_{1n}[eV^2]$ $\varphi_{1n}$ | $I_{2n}[eV]$ $C_{2n}[eV^2]$ $\varphi_{2n}$ | $I_{3n}[eV]$ $C_{3n}[eV^2]$ $\varphi_{3n}$ | $I_{4n}[eV]$ $C_{4n}[eV^2]$ $\varphi_{4n}$ |
|---|---|---|---|---|---|
| $Pb^{0+}$ | 22.45 175.00 2.00 | 33.33(5d) 7.50(5d) 6.00(5d) | 98.39(5p) 2.00(5p) -1.00(5p) | 137.72(5s) 2.00(5s) 5.00(5s) | 179.29(4f) 68.00(4f) 5.00(4f) |
| $Pb^{1+}$ | 39.54 72.00 0.50 | 40.78(5d) 0.10(5d) 6.00(5d) | 105.85(5p) 5.00(5p) -0.90(5p) | 145.20(5s) 2.00(5s) 5.00(5s) | 186.77(4f) 5.00(4f) 5.00(4f) |
| $Pb^{2+}$ | 62.34 255.00 0.50 | 49.46(5d) 0.02(5d) 6.00(5d) | 114.55(5p) 0.02(5p) -1.00(5p) | 153.94(5s) 4.00(5s) -0.80(5s) | 195.53(4f) 1.00(4f) -0.80(4f) |
| $Pb^{3+}$ | 111.12 265.00 -0.50 | 60.13(5d) 0.20(5d) 6.00(5d) | 121.56(5p) 10.20(5p) 3.00(5p) | 164.49(5s) 0.01(5s) -0.80(5s) | 206.26(4f) 5.00(4f) 0.80(4f) |
| $Pb^{4+}$ | 151.05 285.00 -0.60 | 136.90(5p) 4.80(5p) 1.00(5p) | 175.91(5s) 0.01(5s) -0.80(5s) | 218.12(4f) 62.00(5d) 6.00(5d) | 494.11(5d) 5.00(4f) 0.80(4f) |
| $Pb^{5+}$ | 194.49 255.00 -0.60 | 154.07(5p) 0.30(5p) 1.00(5p) | 193.72(5s) 3.00(5s) -0.30(5s) | 237.06(4f) 15.20(5d) 1.50(5d) | 513.02(5d) 5.30(4f) 1.00(4f) |
| $Bi^{1+}$ | 41.78 400.00 0.80 | 116.93(5p) 5.00(5p) 5.00(5p) | 157.87(5s) 6.00(5s) 0.90(5s) | 209.90(4f) 2.00(5s) 5.00(5d) | 493.25(5d) 255.00(4f) 5.00(4f) |
| $Bi^{2+}$ | 70.86 325.00 0.20 | 125.99(5p) 2.40(5p) -0.80(5p) | 166.96(5s) 2.00(5s) 1.00(5s) | 219.05(4f) 1.00(5s) 0.20(5d) | 502.40(5d) 16.00(4f) 0.10(4f) |
| $Bi^{3+}$ | 101.30 315.00 -0.50 | 135.99(5p) 1.00(5p) -0.95(5p) | 176.88(5s) 7.00(5s) 1.00(5s) | 229.01(4f) 4.00(5s) 0.10(5d) | 512.53(5d) 10.00(4f) -0.10(4f) |
| $Bi^{10+}$ | 403.20 775.00 -0.65 | 257.37(5p) 0.01(5p) 5.00(5p) | 301.16(5s) 0.05(5s) 5.00(5s) | 364.14(4f) 0.05(5s) 0.20(5d) | 647.31(5d) 3.20(4f) 0.00(4f) |
| $U^{0+}$ | 17.33 25.00 2.00 | 28.19(6p) 1.00(6p) -0.70(6p) | 45.79(6s) 4.50(6s) 0.40(6s) | 120.92(5d) 12.00(5d) -0.80(5d) | 218.65(5p) 10.50(5p) -0.90(5p) |
| $U^{10+}$ | 303.08 490.00 -0.50 | 258.94(5d) 0.05(5d) 5.00(5d) | 358.96(5p) 2.70(5p) -0.70(5p) | 415.13(5s) 8.00(5s) 2.00(5s) | 592.44(4f) 25.50(4f) 5.00(4f) |
| $U^{13+}$ | 537.35 690.00 -0.80 | 998.04(4d) 20.50(4d) 1.00(5p) | 412.32(5p) 0.05(5p) 0.00(5p) | 468.45(5s) 0.05(5s) 2.00(5s) | 648.51(4f) 7.50(4f) 0.00(4f) |